\documentstyle{article}
\begin{document}
\author{Anatoly Konechny and  Albert Schwarz\\ 
 \\
Department of mathematics, University of California \\
Davis, CA 95616 USA\\
konechny@math.ucdavis.edu,  schwarz@math.ucdavis.edu}
\title{\bf BPS states on noncommutative tori and duality}
\date{December 9, 1998}
\maketitle
\begin{abstract} 
We study gauge theories on noncommutative tori. It was 
proved in \cite{ASMorita} that Morita equivalence of 
noncommutative tori leads to a physical equivalence 
($SO(d,d|{\rm \bf Z})$-duality) of the corresponding gauge theories. 
We calculate the energy spectrum of maximally supersymmetric BPS 
states in these theories and show that this spectrum agrees with 
the $SO(d,d|{\rm \bf Z})$-duality. The relation of our results with those 
of recent calculations is discussed. 
\end{abstract}
\large
\section{Introduction}
It is shown in \cite{CDS} that supersymmetric Yang-Mills (SYM) 
theories on noncommutative tori arise very naturally as compactifications 
of M-theory (in the Matrix formulation of this theory). They can be 
interpreted as toroidal compactifications with nonvanishing 
expectation value of antisymmetric $B$-field in string theory language. 
It was proven later in \cite{ASMorita} that the mathematical notion of 
Morita equivalence is closely related to duality in physics. The results 
of \cite{ASRieffel} and \cite{ASMorita} show that Morita equivalence 
of $d$-dimensional noncommutative tori  is governed by the group 
$SO(d,d|{\rm \bf Z})$. Yang-Mills theories on noncommutative tori 
$T_{\theta}$ and $T_{\hat \theta}$ where $\theta$ and $\hat \theta$ 
are $d\times d$ antisymmetric matrices belonging to the same orbit 
of the group $SO(d,d|{\rm \bf Z})$, are physically equivalent. 
This equivalence is related to T-duality in string theory. The equivalence 
of different compactifications was proven in \cite{ASMorita} at the 
level of action functionals. Of course, this result implies the 
coincidence of energy spectra of the corresponding quantum theories. 
It is necessary to emphasize that this result does not depend on 
the specific form of  action functional and can be applied to any 
gauge-invariant action functional expressed in terms of gauge fields and 
fields in adjoint representation of the gauge group (endomorphisms). 
In particular, it can be applied to Born-Infeld action functional. 
The energy spectrum cannot be expressed in terms of the action functional 
in a simple way. However, in supersymmetric case one can analyze the 
energies of BPS states. We check in this paper that BPS spectra of 
 YM theories on Morita equivalent tori coincide. This coincidence was
 conjectured at first in \cite{CDS} and analyzed later in a number of 
papers (\cite{Ho}, \cite{HofVer1}, \cite{HofVer2}, \cite{BrMor}).
The paper \cite{HofVer2} contains an important observation that quantum 
fluctuations of BPS fields were not taken into account in \cite[v1]{BrMor} 
and that the proper treatment of fluctuations could essentially change the 
answer. It follows from the results of our paper that, indeed, the 
contribution of fluctuations cannot be separated from the contribution 
of ``Kaluza-Klein modes'' (such a separation was assumed in \cite{Ho}
and implicitly in \cite[v1]{BrMor}).
However, we will see that in the case when transverse oscillators are 
in the ground state the energy spectrum agrees with 
\cite[v1]{BrMor} and with 
the results 
of our previous calculation mentioned in \cite[v1]{BrMor}.
 We will see also that our results agree with the results 
of \cite{HofVer2} and \cite[v2]{BrMor} if one takes into account the 
energy of quantum fluctuations. (The quantum numbers used in these papers 
are not independent if all transverse oscillators are in the ground state.)

The paper is  organized as follows. We start with an explanation of 
some basic notions of noncommutative geometry and of the results proved 
in \cite{ASMorita}. Our exposition is different from the one given in 
\cite{ASMorita} in one important relation. The paper  \cite{ASMorita} 
was based on the theory of $C^{*}$-algebras and $C^{*}$-modules. 
In the present paper we give the main definitions and prove that 
complete Morita equivalence leads to a physical equivalence of YM action
 functionals in a more general framework of arbitrary associative algebras. 
It seems that this modification essentially reduces the amount of 
mathematical information needed for the understanding of the relation 
between Morita equivalence and duality. 
We give a detailed formulation 
of the results of \cite{ASRieffel} and \cite{ASMorita} about Morita 
equivalence of noncommutative tori but do not give new proofs of them 
(see, however, Appendix C). 
In this relation we would like to attract attention of the reader to 
the paper \cite{BrMorZum}. Along with other results it contains a new 
proof of the results of \cite{ASRieffel} and of some of the results 
of \cite{ASMorita} that should be more acceptable for a reader that 
finds the exposition in \cite{ASRieffel} and \cite{ASMorita} too 
mathematical. To set a correspondence between the mathematical terminology 
used in \cite{ASRieffel}, \cite{ASMorita} and the one used in 
\cite{BrMorZum} one should notice that ``adjoint sections on twisted 
bundles'' of \cite{BrMorZum} are ``endomorphisms of modules'' in the 
terminology of \cite{ASRieffel}, \cite{ASMorita} and the calculation 
of the ``space of adjoint sections'' performed in \cite{BrMorZum} leads to a 
description of Morita equivalent torus.

In section 3 we give a semiclassical calculation of the energies 
of the BPS states. We start with the detailed analysis of the case of 
two-dimensional noncommutative torus. This analysis permits us 
to verify the relation between energy spectra of BPS states on 
Morita equivalent tori that follows from the results of \cite{ASMorita}. 
At the end we analyze the energies of BPS states in the $d$-dimensional 
case using the results of \cite{ASMorita}. 
(In this paper we consider only maximally supersymmetric BPS states. 
More precisely, we are studying states that arise from quantization of 
maximally supersymmetric BPS fields and fluctuations of these fields. 
Other BPS states on noncommutative tori will be studied  in a forthcoming 
paper \cite{KS}.) Some information about geometric quantization 
that is useful (but not necessary) for understanding of our calculations 
is relegated to  Appendix A. Information about spinor representation of 
$SO(d,d|{\rm \bf Z})$ is collected in Appendix B. Using this information 
we modify proofs given in \cite{ASMorita} keeping track of all constant 
factors (Appendix C). Appendix D contains a description 
of modules over noncommutative tori that can be equipped with a constant 
curvature connection.


\section{General Theory}
Let us consider an associative algebra $A$. We can interpret it as an algebra of functions on a 
``noncommutative space '' and introduce various geometric notions generalizing  notions of
 standard ``commutative'' geometry. In particular, we can define the notion of a connection on 
$A$-module $E$. Recall that by definition a 
linear space $E$ is a left $A$-module if we can multiply elements of $E$ by elements of 
$A$ from the left and $ a(be)=(ab)e$ , $a(e+e') = ae + ae' $ , $a(\lambda e)= \lambda (ae) $ 
(here $a,b \in A ,  \enspace e, e' \in E$, $\lambda$ is 
a number). The definition of a right $A$-module is similar. 
Direct sum $A^{n}$ of $n$ copies of $A$ can be considered both as 
a left and a right module over $A$ in a natural way. 
Such a module is called a free module. If $A$ is a commutative algebra 
$C(M)$ (or $C^{\infty}(M)$) of continuous (or smooth) functions on a compact 
manifold $M$, we can consider the space $\Gamma(E)$ of (continuous or smooth) 
sections of vector bundle $E$ over $M$ as a module over $A$. 
(The distinction between left and right modules disappears for commutative 
algebras.) One can check that $A$-modules obtained by means of this construction 
can be characterized as finitely generated projective modules 
(i.e. direct summands in free modules).

We will give a definition of a connection on an $A$-module $E$,
taking as a starting point a Lie algebra $L$ that acts on $A$ by means of infinitesimal automorphisms 
(derivations). In other words we assume that we fixed operators $\delta_{X}$ depending linearly on 
$X\in L$ and obeying the identity 
$\delta_{X} (ab) = (\delta_{X} a )b + a (\delta_{X} b)$. 
Then a connection on $E$ is specified by means of linear operators 
$\nabla_{X} : E \to E \enspace , X \in G$, obeying the Leibnitz rule: 
$$ 
\nabla_{X} (a e) = a \nabla_{X} e + (\delta_{X} a ) e 
$$
for any $a\in A$ , $e\in E$. (We formulated the definition in the case of a left $A$-module $E$. 
The definition for a right $A$-module is similar.) There exists a more general definition of  
connection where covariant derivatives $\nabla_{X}$ are replaced by covariant  differentials; 
we do not use this notion. A curvature of a connection $\nabla_{X}$ can be defined by the formula 
$$
F_{XY} = [\nabla_{X} , \nabla_{Y} ] - \nabla_{[X,Y]} \, . 
$$
It is easy to check that  $aF_{X,Y}=F_{X,Y}a$  for any $a \in A$ , $X,Y\in L$. 
A linear operator $\phi : E \to E$ is called an endomorphism of an $A$-module $E$ if 
it is $A$-linear, i.e. it commutes with multiplication by elements of $A$. We see that 
a curvature of a connection can be considered as a two-form on the Lie algebra $L$ 
that takes values in the algebra $End_{A}E$ of endomorphisms of the $A$-module $E$. 
We will restrict ourselves to the case when the Lie algebra $L$ is abelian. 
Then the second 
term in the expression for curvature tensor vanishes. Let us assume that the algebra 
$End_{A}E$ is equipped with a trace (i.e. with a linear functional $\rm Tr$ obeying 
${\rm Tr} \alpha \beta = {\rm Tr} \beta \alpha$ ). Then we can construct a Yang-Mills action functional 
$S^{YM}$ on the set $Conn$ of connections in the $A$-module $E$ 
by means of an inner product on the Lie algebra $L$. Namely, we can use the formula 
$$
S^{YM}(\nabla ) = \frac{1}{g_{YM}^{2}} \sum_{\alpha, \beta} {\rm Tr} 
F_{\alpha \beta}F^{\alpha \beta}  
$$
Here $F_{\alpha \beta}$ stands for components of curvature tensor in some basis of $L$ and 
$F^{\alpha \beta} = g^{\alpha \mu} g^{\beta \nu} F_{\mu \nu}$ where $g^{\alpha \mu}$ is the 
inverse metric tensor on $L$ . If $dim L = 10, 6 , 4 , 3$ one can construct supersymmetric 
extension of $S^{YM}$ in the usual way; we do not need the explicit form of this extension. 
We can also consider the Yang-Mills action in the presence of a background field $\phi$ : 
$$
S^{YM}_{\phi}(\nabla) = 
\frac{1}{g_{YM}^{2}} \sum_{\alpha \beta} {\rm Tr} (F_{\alpha \beta}  + \phi_{\alpha \beta} \cdot 1 ) 
(F^{\alpha \beta} + \phi^{\alpha \beta} \cdot 1) . 
$$

The most important example of a ``noncommutative space '' is a noncommutative torus. 
The algebra $A_{\theta}$ of ``smooth functions'' on a noncommutative torus can be 
defined as a linear space  $S(Z^{d})$ equipped with multiplication given by the formula 
\begin{equation} \label{mult}
(f * g)(\gamma) = \sum_{\lambda \in Z^{d}} e^{\pi i \theta_{\lambda, \gamma - \lambda} } 
f(\lambda)g(\gamma - \lambda). 
\end{equation}
Here $\theta$ is an antisymmetric bilinear form on $Z^{d}$ and $S(Z^{d})$ stands 
for Schwartz space (the space of complex functions on $Z^{d}$ that decrease faster than 
any power function). Instead of the antisymmetric bilinear form we consider the
antisymmetric $d \times d$ matrix of its coefficients and use the same notation $\theta$ for it. 
If $D=Z^{d}$ is considered as a lattice in $R^{d}$ , then the dual space $R^{*d}$ is an 
abelian group that acts naturally on $S(Z^{d})$ : to every $x\in R^{*d}$ we assign a map 
$\tau_{x}$ transforming a function $f(\lambda)$ into $e^{2\pi i <x, \lambda > } f(\lambda)$. 
For every $\theta$ the map $\tau_{x}$ can be considered as an automorphism of $A_{\theta}$ ; 
this automorphism is trivial if $x\in D^{*}$ where $D^{*}$ is the lattice dual to $D$. 
We denote the group $R^{*d}/D^{*}$ considered as a group of automorphisms of $A_{\theta}$ 
by $\tilde L_{\theta}$ and its Lie algebra as $L_{\theta}$. 
If $\theta = 0$ (or, more generally, the matrix $\theta$ has integer 
entries ) , then the algebra $A_{\theta}$ is isomorphic to the algebra of smooth functions on the standard (``commutative'') torus. It is easy to check that the notion of connection 
and the expression for Yang-Mills action functional coincide with the standard ones 
in this  case. 
We will define a connection on $T_{\theta}$-module by means of an isomorphism 
$L\to L_{\theta}$. Using this isomorphism and a standard basis in the lattice 
$D^{*}\subset L_{\theta}$ we obtain a standard basis in $L$. 
It consists of elements   $X^{i}$ such that 
\begin{equation} \label{stbasis}
(\delta_{X_{j}}f)(n^{i}e_{i}) \equiv (\delta_{j}f)(n^{i}e_{i}) = 
i n^{j}f(n^{i}e_{j})  \, , j = 1, \dots , d \, . 
\end{equation}

Let us come back to the general case and define the notions of Morita equivalence and
 complete Morita   equivalence of associative algebras. Consider an $(A, \hat A)$-bimodule 
$P$ (i.e. we assume that elements of $P$ can be multiplied by elements of $A$ from the left and 
elements of $\hat A$ from the right in such a way that $(a e)\hat a = a (e \hat a)$ for any 
$a\in A$, $\hat a \in \hat A$, $e\in P$, and $P$ with this operations 
can be considered   as a left $A$-module and a right $\hat A$-module). Notice that 
every left $A$-module $E$ can be regarded as an $(A, End_{A} E)$-bimodule and conversely, 
for every $(A, \hat A)$-module $P$ there exists a natural map $\hat A \to End_{A} P$. 
For every right $A$-module $E$ and $(A, \hat A)$-bimodule $P$ we can construct 
a right $\hat A$-module $\hat E = E\otimes_{A} P$. (To define the tensor product over 
$A$ we identify $ea\otimes p$ with $e\otimes ap $ in standard tensor product of linear 
spaces $E$ and $P$. This identification respects multiplication by $\hat a\in \hat A$ ,  
therefore, $\hat E$ can be considered as a right $\hat A$-module.) For every $A$-linear 
map of $A$-modules $\alpha : E\to E'$ we can construct naturally an $\hat A$-linear 
map $\hat \alpha: \hat E \to \hat E' $. It is easy to check that 
$\widehat{\alpha \beta} = \hat \alpha \hat \beta $. In particular, we obtain a map from 
$End_{A} E$ into $End_{\hat A} \hat E$. We say that an $(A, \hat A)$-bimodule $P$ is 
an equivalence bimodule if there exists an $(\hat A, A)$-module $P'$ obeying 
$P\otimes_{\hat A}P'=A$ and $P'\otimes_{A}P=\hat A$. (Here $A$ and $\hat A$ are considered 
as $(A,A)$-bimodule and $(\hat A, \hat A)$-bimodule respectively.) Algebras $A$ and $\hat A$ 
are called Morita equivalent if there exists an equivalence $(A,\hat A)$-bimodule $P$. 
In mathematical terms every $(A, \hat A)$-bimodule $P$ determines a functor $E\to \hat E$ 
acting from the category of right $A$-modules into the category of right $\hat A$-modules. 
The $(A, \hat A)$-bimodule $P$    specifies Morita equivalence of algebras $A$ and $\hat A$ 
if this functor is an equivalence of categories of modules. (It is easy to derive from our assumptions 
about  $P'$, that $\hat E\otimes_{\hat A}P' = E$ and for every right $\hat A$-module $F$ we 
have $(F\otimes_{\hat A} P')\otimes_{A}P=F$. This means that the existence of $P'$ implies 
equivalence of categories of modules; the inverse statement is also true.)

Let us assume now that $(A, \hat A)$-bimodule $P$ is equipped with operators $\nabla_{X}^{P}$ 
that satisfy 
$$ 
\nabla_{X}^{P}(ae)=a\nabla_{X}^{P}e + (\delta_{X}a) e 
$$
$$
\nabla_{X}^{P} (e\hat a) = (\nabla_{X}^{P}e) \hat a + e\tilde \delta_{ X}\hat a 
$$
$$
[\nabla_{X}^{P}, \nabla_{Y}^{P} ] = \sigma_{XY} \cdot 1 \enspace . 
$$
 Here $\nabla_{X}^{P}$ is a linear operator in $P$ that depends linearly on an element $X$ of 
abelian Lie algebra $L$, $a\in A$, $\hat a \in \hat A$, $e\in P$ ; the Lie algebra $L$ acts 
on $A$ and $\hat A$ by means of operators $\delta_{X}$ and $\tilde \delta_{X}$ correspondingly. 
 The above conditions mean that $\nabla_{X}^{P}$ is a connection in $P$ considered as a left 
$A$-module or as a right $\hat A$-module, and that the curvature of this connection is constant. 
Using the operators $\nabla_{X}^{P}$ we can construct a connection $\hat \nabla_{X}$ in a 
$\hat A$-module $\hat E=E\otimes_{A}P$ for every connection $\nabla_{X}$ in an $A$-module $E$. 
This construction is based on a remark that the operator $\nabla_{X}\otimes 1 + 1 \otimes \nabla_{X}^{P}$ 
in $E\otimes P$ descends to $\hat E=E\otimes_{A} P$ and determines a connection $\hat \nabla_{X}$ 
in $\hat E$. One can check that the curvature of the connection $\hat \nabla_{X}$ can be expressed 
in terms of the curvature $F_{XY}^{\nabla}$ of the connection $\nabla_{X}$: 
\begin{equation} \label{F}
F_{XY}^{\hat \nabla} = \hat F^{\nabla}_{XY} + \sigma_{XY}\cdot 1 \, . 
\end{equation}
(Recall that for $X,Y \in L$ we consider $F_{XY}^{\nabla}$ as an element of $End_{A} E$ and 
therefore $\hat F_{XY}^{\nabla}$ is defined as an element of $End_{\hat A} \hat E$.)

Now we can define a complete Morita equivalence of algebras $A$ and $\hat A$ by means 
of $(A,\hat A)$-bimodule $P$ equipped with connection $\nabla_{X}^{P}$.  Namely, we 
assume that  $P$ is an equivalence bimodule and that the above correspondence of 
connections in $E$ and $\hat E$ is bijective. If such a bimodule $P$ exists we say that $A$ 
and $\hat A$ are completely Morita equivalent. It is clear that complete Morita equivalence 
of algebras $A$ and $\hat A$ implies physical equivalence of Yang-Mills theories on 
$A$-module $E$ and corresponding to it $\hat A$-module $\hat E$. More precisely, for a 
trace ${\rm Tr}$ on $End_{A} E$ we define a trace $\widehat{\rm Tr}$ on $End_{\hat A}\hat E$ by the 
formula 
\begin{equation} \label{traces}
\widehat{\rm Tr}\hat \alpha = {\rm Tr} \alpha \, . 
\end{equation} 
Then 
\begin{equation} \label{actions} 
S^{YM}_{\phi} ( \nabla) = S^{YM}_{\phi - \sigma}(\hat \nabla) \, . 
\end{equation}

We will consider the case when $E$ is a finitely generated projective module. 
Then for every trace on $A$ one constructs a trace on $End_{A}E$. If the trace 
on $A$ is normalized (i.e. ${\rm Tr} 1 = 1$), the trace on $End_{A}E$ is not 
necessarily normalized. The dimension $dimE$ of module $E$ can be defined as the value 
of the trace on the identity endomorphism: ${\rm Tr} {\bf 1} = dimE$.
 If we use this trace we should modify (\ref{actions}) (see below).

In the most interesting cases the algebras at hand are equipped with an involution
 (complex conjugation). 
Moreover, these algebras are $C^{*}$-algebras and modules over them are  
 $C^{*}$-modules . In these cases it is natural to restrict the attention  to Hermitian 
connections. A theory of Morita equivalence  of $C^{*}$-algebras was developed 
by Rieffel \cite{Rieffel}; the consideration in \cite{ASMorita} was based on it. 
In some relations 
this theory is easier to apply (for example the $(\hat A, A)$-bimodule $P'$ 
in the definition of Morita equivalence can be obtained from $P$ by means of 
complex conjugation).

To apply the equivalence (\ref{actions}) in concrete situations we should find the $\hat A$-module 
$\hat E$ corresponding to given $A$-module $E$. We can use Chern characters to 
solve this problem. 
We consider the situation when the connections on $A$-module $E$ are defined 
by means of action of Lie algebra $L$. Then the Chern character $ch(E)$ can be defined 
by the formula 
\begin{equation} \label{Chern}
ch(E) ={\rm Tr}\,  exp\left( \frac{1}{2\pi i} \alpha^{k} F_{kj}^{\nabla} \alpha^{j} \right) 
= \sum_{k=0}^{\infty} \frac{1}{(2\pi i)^{k}k!} {\rm Tr}(F^{\nabla})^{k}   \, . 
\end{equation}
Here $F^{\nabla}=\alpha^{k}F_{kj}^{\nabla}\alpha^{j} $ stands for  the curvature of connection 
$\nabla$ on $E$ considered as a two-form on $L$ with values in $End_{A} E$ or as an 
element of $\Lambda^{n}\otimes End_{A} E $ where $\Lambda^{d}$ is a Grassmann algebra 
with generators $\alpha^{1}, \dots \alpha^{d} , d=dim L$. Everywhere below for shortness
 we  drop coefficients $1/2\pi i$ from the formulas involving Chern characters. 
The Chern character $ch(E)$ is 
an element of $\Lambda^{n}=\Lambda(L^{*})$ (an inhomogeneous form on $L$). It does 
not depend on the choice of  connection $\nabla$.

If the algebras $A$ and $\hat A$ are completely Morita equivalent one can relate 
the Chern characters of  $E$ and $\hat E$ using (\ref{F}). 
We obtain\footnote{Note that in the corresponding formula in \cite{ASMorita} (formula (41)) 
as well as in other formulas of Sec. 5 one should replace $\phi$ with $\sigma$.}  
\begin{equation} \label{ch}
ch(\hat E) = e^{\alpha^{k}\sigma_{kj}\alpha^{j}} ch(E) \, . 
\end{equation}
We should slightly change this formula if the trace on $End_{A}E$ is normalized by 
the condition ${\rm Tr} {\bf 1} = dimE$. Namely, we have 
\begin{equation} \label{8_1/2}
ch(\hat E)=\frac{dim \hat E}{dim E}  
e^{\alpha^{k}\sigma_{kj}\alpha^{j}} ch(E) \, . 
\end{equation}
Now we restrict ourselves to the case when $E$ is a finitely generated projective module 
(direct summand in a free module) and the algebra $A$ is the noncommutative torus 
$T_{\theta}$. 
 Let us mention first of all that there is 
a canonical trace on $T_{\theta}$ given by the formula ${\rm Tr} ( f ) = f( 0 )$ 
if $f\in T_{\theta}$ is considered as a function on a lattice. If $\theta$ is 
irrational (i.e. has at least one irrational entry) this trace is unique up to 
a constant factor. Similar statement is correct for the algebra 
$End_{T_{\theta}}E$ provided $\theta$ is irrational. On this algebra we 
normalize the trace by the condition ${\rm Tr} 1 = dim E$.  
In what follows we always work with  this trace. 
For the case of a noncommutative torus $T_{\theta}$  one can prove \cite{Elliott} that the expression 
\begin{equation} \label{chmu}
\mu (E) = e^{-\frac{1}{2} b_{k} \theta^{kj} b_{j}} ch(E) 
\end{equation}
 where $b_{k}$ stands for the derivative with respect to anticommuting variable 
$\alpha^{k}$, is an integral element of the Grassmann algebra $\Lambda = \Lambda(L^{*}) $, 
i.e. an element of $\Lambda(D)$. (Recall that the group $\tilde L_{\theta}$ of automorphisms 
of $T_{\theta}$ can be represented as $L_{\theta}/D^{*} = R^{*d}/D^{*}$. 
One can identify $\Lambda(L^{*})$ with cohomology algebra of the torus $\tilde L_{\theta}$ and  
$\Lambda(D)$ with integer cohomology $H(\tilde L_{\theta}, Z)$.) 
The integer coefficients in the representation of $\mu(E)$ in the basis of $D$ play the role of 
topological numbers of the module $E$.

Notice that the Grassmann algebras $\Lambda(L^{*}) = F^{*}$ and $\Lambda(L)=F$
can be considered as fermionic Fock spaces (irreducible representations of a 
finite-dimensional Clifford algebra). In particular, in $F^{*}=\Lambda(L^{*})$ we 
have operators $a^{k}$ of multiplication by $\alpha^{k}$ and operators 
$b_{k}=\frac{\partial}{\partial \alpha^{k}}$ satisfying canonical  anticommutation 
relations 
\begin{equation} \label{CAR}
\{a^{k}, b_{l} \}_{+} = \delta_{l}^{k} \, , \enspace 
\{a^{k}, a^{l} \}_{+} = 0 \, , \enspace 
\{b_{k}, b_{l} \}_{+} = 0 
\end{equation}
The group $O(d,d|{\rm \bf C})$ can be regarded as a group of automorphisms 
of Clifford algebra (a group of linear canonical transformations).  
Operators given by the formulas 
$$
\tilde a^{k} = M_{l}^{k} a^{l} - N^{kl}b_{l} \, , \enspace 
\tilde b_{k} = -R_{kl}a^{l} + S_{k}^{l}b_{l} 
$$
obey canonical anticommutation relations (\ref{CAR})  iff the matrix 
\begin{equation} \label{g1}
g= \left( \begin{array}{cc}
M&N\\
R&S\\
\end{array} \right) 
\end{equation}
belongs to the group $O(d,d|{\rm \bf C})$. 
Using this remark one can define a projective action of $O(d,d|{\rm \bf C})$ on 
$F^{*}$ assigning to every $g\in O(d,d|{\rm \bf C})$ an operator $V_{g}: F^{*} \to F^{*}$ 
that satisfies 
\begin{equation} \label{Vg}
\tilde a^{k} = V_{g}a^{k}V_{g}^{-1} \, , \enspace 
\tilde b_{k} = V_{g}b_{k}V_{g}^{-1} \, . 
\end{equation}
The projective action is (by definition) the spinor representation 
of $O(d,d|{\rm \bf C})$.
We also define an action  $\theta \mapsto g\theta = \hat \theta$   of $O(d,d|{\rm \bf C})$ on 
the space of  antisymmetric matrices by the formula 
\begin{equation} \label{newteta}
\hat \theta = (M\theta + N)(R\theta + S)^{-1}
\end{equation}
where $d\times d$ matrices $M$, $N$, $R$, $S$ correspond to an element $g\in O(d,d|{\rm \bf C})$ 
by formula (\ref{g1}).
More precisely, this action is defined on a subset of the space of all antisymmetric matrices 
where the matrix $R\theta + S$ is invertible.  
The main results of \cite{ASMorita} can now be formulated in the following way. 


{\it Tori $T_{\theta}$ and $T_{\hat \theta}$ are completely Morita equivalent iff 
$\hat \theta$ and $\theta$ are related by the formula  (\ref{newteta}) 
where the matrix $g$ defined by the formula (\ref{g1})
belongs to the subgroup $SO(d,d|{\rm \bf Z})$ of $SO(d,d|{\rm \bf C})$ consisting of matrices 
with integer entries. } \\
(One can say that $\theta$ and $\hat \theta$ should belong to the same 
orbit of $SO(d,d|{\rm \bf Z})$.)

{\it If $E$ is a $T_{\theta}$-module, $\hat E$ is the corresponding 
$T_{\hat \theta}$-module, then 
\begin{equation}\label{mues}
\mu(\hat E) = V_{g}\mu(E) \, .
\end{equation} 
}

Given two Morita equivalent tori $T_{\theta}$ 
and $T_{\hat \theta}$, the connections on modules $E$ and $\hat E$ are defined 
for fixed isomorphisms: 
$$
\delta : L \to L_{\theta}
$$
$$
\tilde \delta : L \to L_{\hat \theta} \, . 
$$
These isomorphisms determine two standard bases (\ref{stbasis}) in $L$ that we 
denote as $X^{i}$ and $\tilde X^{i}$. 
It is proved in \cite{ASMorita} (formulas (45) and (55)) that 
$\tilde X^{i} = A^{i}_{j} X^{j}$ where the matrix $A$ 
can be expressed as 
\begin{equation} \label{A} 
A = S + R\theta \, . 
\end{equation}
(See also Appendix C for the proof of this formula and formulas (\ref{mues}) and 
(\ref{dimensions}).) 
 The metric tensors in different standard bases are related by the formula 
\begin{equation} 
\hat g_{ij} = A^{k}_{i}g_{kl}A^{l}_{j} \, .
\end{equation} 
Therefore, given the transformation from $SO(d,d|{\rm \bf Z})$ that 
relates Morita-equivalent tori 
$T_{\theta}$ and $T_{\hat \theta}$ we can find the metric $\hat g_{\alpha \beta}$. 
To find $\hat F$ we may use the expression (\ref{F}). When written in the basis 
$\tilde X^{i}$ it reads as 
\begin{equation}\label{curvatures}
A_{i}^{k}F_{kl}A^{l}_{j}  +  \sigma_{ij} =  F^{\hat \nabla}_{ij} \, . 
\end{equation}
As shown in \cite{ASMorita} (formula (55)) the matrix $\sigma$ is given by the expression 
\begin{equation} \label{sigma}
\sigma =-RA^{t}=-R(S +  R\theta)^{t} \, . 
\end{equation}
Finally, the dimension $dim \hat E$ can be calculated using formula (\ref{ch}): 
\begin{equation} \label{dimensions}
dim\hat E = dim E |det(S + R\theta )|^{-1/2} \, . 
\end{equation}
The relation (\ref{dimensions}) was given in \cite{BrMorZum}. It can be obtained 
from the results of \cite{ASMorita} if we take into account that $dim E$ is equal 
to the value of Chern character $ch(E)$ at the point $\alpha=0$. 
One should use the formula
\begin{equation} \label{V3}
ch(\hat E)=e^{\alpha^{k}\sigma_{kj}\alpha^{j}}V_{3} ch(E) 
\end{equation} 
where $V_{3}$ is a canonical transformation having the form 
$$ 
(V_{3}f)(\alpha ) = c\cdot f((A^{t})^{-1} \alpha ) \, .
$$
One can check that the operator $V_{3}$ in (\ref{V3}) preserves an appropriate 
bilinear form in Fock space. Calculating the constant $c$ from this condition 
we obtain 
$$
c = |det A|^{-1/2} = |det(S + R\theta )|^{-1/2} \, . 
$$

We mentioned already  that Yang-Mills theories on modules $E$ and $\hat E$ 
are equivalent. Now we are able to give more detailed description of this 
equivalence in the case at hand.
The formula (\ref{actions}) is derived for the case when the trace on $End_{\hat A}\hat E$
is defined by the formula (\ref{traces}). If we use the canonical normalized trace on 
$End_{T_{\theta}}E$ and $End_{T_{\hat \theta}}\hat E$ we obtain instead the 
following relation 
\begin{equation} \label{actions2}
(dimE)^{-1}S^{YM}_{\phi}(\nabla) = (dim\hat E)^{-1}S^{YM}_{\phi-\sigma }(\hat \nabla) \, .
\end{equation}

We have shown  how one can calculate  the quantities related with the module
$\hat E$ 
(curvature $\hat F$, metric $\hat g_{ij}$ , $dim\hat E$, etc.  ) out of the original ones 
defined on $E$. 
Now we would like to write down the relations above for the particular case of noncommutative 
1+2-tori and Morita equivalence given by a particular $SO(3,3|Z)$ transformation 
to be specified below. We assume that in the standard basis the matrix $\theta_{i,j}$ that 
defines the torus $T_{\theta}$ has the form 
\begin{equation} \label{3dteta}
\theta = \left(\begin{array}{ccc}
0&0&0\\
0&0&\vartheta\\
0&-\vartheta&0\\
\end{array} \right) \, . 
\end{equation}
We say that we have a 1+2- torus meaning that  we fixed the  splitting into $S^{1}$ part and 
the  part isomorphic  to a two-dimensional noncommutative torus. 
We will consider a $SO(3,3|Z)$ transformation that preserves the form (\ref{3dteta}) 
and maps the parameter $\vartheta$ into $-1/\vartheta$. For this transformation 
the matrix  (\ref{g1}) has (matrix) entries $M=0$, $R=I$, 
$$
N=\left( \begin{array}{ccc}
0&0&0\\
0&1&0\\
0&0&1\\
\end{array} \right) \, ,
$$
$R=N$, and $S=I-N$ (where $I$ is the identity matrix.
Modules over 1+2 (noncommutative)  tori are characterized by 
an integral element $\mu(E)=p + 
\frac{1}{2}q_{ij}\alpha^{i}\alpha^{j}$ of the Grassmann algebra $\Lambda(L^{*})$. 
We introduce the notation $E(p;q_{ij})$ for a module having these 
topological numbers. 
Using (\ref{mues}) one can calculate $\mu(\hat E)$ for the
            transformation at hand: 
$\mu (\hat E)= -q_{23} +p\alpha^{2}\alpha^{3} + q_{13}\alpha^{1}\alpha^{2} - 
q_{12}\alpha^{1}\alpha^{3} $. Thus, the Yang-Mills theory on the 
modules $E=E(p;q_{23}, q_{12}, q_{13})$ 
and $\hat E = E(-q_{23};p,q_{13},-q_{12})$ are equivalent. 
Assuming that  $(g_{ij})=diag(R_{0}^{2}, R_{1}^{2}, R_{2}^{2})$ 
(i.e. the metric on 
$T_{\theta}$ is diagonal with the specified entries), one can easily calculate  
$(\hat g_{i,j})=diag(R_{0}^{2}, R_{2}^{2}\vartheta^{2}, R_{1}^{2}\vartheta^{2})$. 
For the curvature $\hat \nabla$ on the module $\hat E$ one obtains the expression 
\begin{equation}\label{1}
F^{\hat \nabla}= 
\left(\begin{array}{ccc}
0& f_{13}\vartheta &-f_{12}\vartheta \\
-f_{13}\vartheta & 0 &f_{23}\vartheta^{2} \\
f_{12}\vartheta & -f_{23}\vartheta & 0 \end{array}   \right) + 
\left( \begin{array}{ccc}
1&0&0\\
0&0&\vartheta \\
0&-\vartheta&0 \end{array} \right) \, .
\end{equation}
For a connection of  constant curvature $f_{ij} = \frac{q_{ij}}{dimE}$ 
one can check that formula (\ref{1}) gives the correct expression 
for the curvature of the constant curvature connection on $\hat E$.  
To compare the dimensions $dimE$ and $dim\hat E$ one can calculate 
$ch(E)$ and $ch\hat E$ using (\ref{chmu}). Comparing the free terms of those expressions 
one gets $dim \hat E = \vartheta^{-1}dimE$. As one can easily check, 
 formula (\ref{actions2}) holds when  explicit expressions for quantities on $\hat E$ are
  substituted to the RHS of the formula.

Let us emphasize that the considerations above can be applied not 
only to the standard YM  action, but also to SUSY YM action, to 
Born-Infeld action, etc. (We can consider any gauge-invariant 
action functionals 
that depend on connections and endomorphisms on a $T_{\theta}$-module 
$E$. In the commutative case this means that we consider gauge fields 
and fields that transform according to the adjoint representation of the gauge group.)


\section{Energy of BPS states} 
The equivalence of SYM action functionals on Morita equivalent tori implies 
the coincidence of the corresponding energy spectra. In this 
section we will explicitly calculate the energies of BPS states of 
the   SYM theory on a noncommutative torus  and 
show that they are invariant under Morita equivalence. 
Our calculation will be semiclassical, but the presence of supersymmetry 
makes the calculation exact for BPS states. Instead of working with the 
full supersymmetric action we will consider  the $1+d$ YM action 
functional and constant curvature solutions which ,by abuse of 
terminology, we will call  BPS fields (they satisfy BPS condition 
in the supersymmetric theory). It is a valid thing to do because the 
calculation  leads to the same result;  
 the sole role of supersymmetry is to ensure the exactness of the 
semiclassical approximation.  
Hence, we start with the following (Euclidean) action functional 
\begin{equation}\label{action} 
S=\frac{1}{4g_{YM}^{2}}{\rm Tr}\left( 
V\sum_{\alpha \beta} 
(F_{\alpha \beta}+ \phi_{\alpha\beta}\cdot {\bf 1})
g^{\alpha \mu} g^{\beta \nu} (F_{\mu \nu}+ \phi_{\mu\nu}\cdot {\bf 1}) 
\right) \, . 
\end{equation}
Here $g_{YM}$ stands for the YM coupling constant, $\phi_{\alpha \beta}$ 
plays the role of a background field,  
$V=R_{0}R_{1}\ldots R_{d}$ is the volume element,   
the indices $\alpha$, $\beta$ take values $0, 1, \dots , d$.

We can fix a connection $\nabla^{0}_{\alpha}$ and 
represent every connection in the form 
$\nabla_{\alpha} = \nabla_{\alpha}^{0} + X_{\alpha}$ where 
$X_{\alpha}$ is an endomorphism. One can represent the functional 
$S$ explicitly as a sum over a lattice. By means of Fourier transform 
we can replace the summation over a lattice with the integration 
over a commutative torus. However, the action functional is nonlocal 
in this representation. We would like to calculate the energies of 
BPS states in Hamiltonian formalism. Therefore, we should single out 
the time direction $x_{0}$ and perform Wick rotation.  To avoid nonlocality 
in time we should assume that $\theta_{0\alpha} = 0$. In other words, we 
suppose that the $(1+ d)$-dimensional noncommutative torus $T_{\theta}$ 
is a direct product of a circle and a $d$-dimensional noncommutative 
torus $T_{\vartheta}$.

In the gauge $\nabla_{0}=\frac{\partial}{\partial t}$ we obtain  
a Hamiltonian  
\begin{equation}\label{hamiltonian}
H= {\rm Tr} \frac{g_{YM}^{2}R_{0}^{2}}{2V}
P^{i}g_{ij}P^{j}  + 
 {\rm Tr}\frac{V}{4g_{YM}^{2}}(F_{ij} + 
\phi_{ij}\cdot {\bf 1})g^{ik}g^{il}(F_{kl} + 
\phi_{kl}\cdot {\bf 1}) \, .  
\end{equation}
Here $\nabla_{i}$ is a connection on a 
$T_{\vartheta}$-module $E$, $P^{i} \in End_{T_{\vartheta}} E$, and 
${\rm Tr}$ denotes the trace in $End_{T_{\vartheta}} E$. Thus, $H$ is 
defined on the space $Conn \times (End_{T_{\vartheta}}E)^{d}$.
In the derivation of (\ref{hamiltonian}) we assumed
 that the metric $g_{\alpha \beta}$ obeys
$g_{0i} = 0$, $g_{00} = R_{0}^{2}$,  
$g_{ij}= \delta_{ij} R_{i}^{2}$ and the antisymmetric  
tensor $\phi_{\alpha \beta}$ has  only spatial nonzero components 
$\phi_{ij}$ (here and everywhere Greek indices $\alpha, \beta, \dots$ 
run from $0$ to $d$ and Latin indices $i, j, \dots$ run from $1$ to 
$d$). 
The Hamiltonian (\ref{hamiltonian}) should be restricted to a 
subspace $\cal N$ where the constraint 
\begin{equation} \label{constr} 
[\nabla_{i}, P^{i}] = 0 
\end{equation}
is satisfied. More precisely, one should consider $H$ as a function 
on the space ${\cal N}/G = {\cal P}$ where $G$ is a group of unitary 
elements of $End_{T_{\vartheta}}E$ (the group of spatial gauge 
transformations). The symplectic form on the space 
$Conn \times (End_{T_{\vartheta}}E)^{d}$ can be written as 
\begin{equation}
\omega = {\rm Tr}\delta P^{i}\wedge \delta \nabla_{i} \, . 
\end{equation}
The restriction of this form to $\cal N$ is degenerate , but it 
descends to a nondegenerate form on ${\cal P} = {\cal N}/G$ 
(on the phase space of our theory). The phase space $\cal P$ is 
not simply connected. Its fundamental group is the group of connected 
components of the gauge group $G$. In other words, 
$\pi_{1}({\cal N}/G) = G/G_{0}\equiv G^{large}$ where $G_{0}$ is the 
group of ``small'' gauge transformations (connected component of $G$). 
One can say that $\pi_{1}({\cal N}/G)$ is a group of ``large'' gauge 
transformations. It is useful to consider the phase space $\cal P$ 
as a quotient $\tilde {\cal P}/G^{large}$ where 
$\tilde {\cal P} = {\cal N}/G_{0}$ is a symplectic manifold obtained 
from $\cal N$ by means of factorization with respect to  small 
gauge transformations.

A Hamiltonian system on a phase space $\cal P$ with a nontrivial 
fundamental group can be quantized in several nonequivalent
 ways (see Appendix A). The freedom 
is labeled by characters of $\pi_{1}({\cal P})$ (or, equivalently, 
by elements of the cohomology group 
$H^{1}({\cal P}, {\rm \bf R/Z})$). 
One can verify that the study of our system in Lagrangian formalism 
in the presence of topological terms in the action can be reduced 
to the analysis of the system in Hamiltonian formalism if all 
possible ways of quantization are taken into account.


Now let us consider the case $d=2$ in full detail. Let us fix 
 a torus $T_{\vartheta}$ corresponding to the matrix 
$$
\left( \begin{array}{cc} 
0&\vartheta\\
-\vartheta & 0\\
\end{array} \right) \, .
$$ 
Modules over $T_{\vartheta}$ 
are labeled by pairs of integers $(p,q)$ obeying $p-q\vartheta > 0$. 
An explicit description of these modules $E_{p,q}$ can be found in 
\cite{CR}, or \cite{CDS}. We do not need it.  Let us mention only 
that $dim E_{p,q} = p - q\vartheta$, 
$\mu(E_{p,q}) = p + q\alpha_{1}\alpha_{2}$, 
$ch(E_{p,q}) = dimE_{p,q} + q\alpha_{1}\alpha_{2}$, and the curvature of a 
constant curvature connection is 
$F_{12}=\frac{q}{dimE_{p,q}}\cdot {\bf 1}$ in the standard basis.
(Note that here and everywhere below  we  omit the $2\pi i$ factor that stands 
at some integers.)  
 Let us consider at first 
modules $E_{p,q}$ where $p$ and $q$ are relatively prime. These 
modules can be called basic modules because every module 
$E_{p,q}$ can be represented as a direct sum of $D$ copies of 
identical basic modules $E_{p', q'}$. (Here $D= g.c.d.(p,q)$ 
and $Dp'=p$, $Dq'=q$.) Let us fix a constant curvature connection 
$\nabla^{0}$. It follows from the results of \cite{CR} that for 
a basic module any other constant curvature connection can be 
transformed to the form $\nabla^{0}_{j} + iq_{j}\cdot {\bf 1}$ 
by means of small gauge transformations. Using large gauge transformations 
one can prove that  $\nabla^{0}_{j} + iq_{j}\cdot {\bf 1}$  is 
gauge equivalent to  
$\nabla_{j}^{0} + i(q_{j} - \frac{ n_{j}}{dimE_{p,q}})\cdot {\bf 1}$    
where $n_{j} \in {\rm \bf Z}$. Therefore, the space of gauge 
classes of constant curvature connections is a two-dimensional torus.
From now on we fix $p$, $q$ and omit subscripts in the notation of the module 
$E_{p,q}$.

We will say that a set $(\nabla_{i}, P^{j})$ is a BPS field if 
$\nabla_{i}$ is a constant curvature connection and 
$P^{j} = p^{j}\cdot {\bf 1}$. (This terminology is prompted by 
the fact that after supersymmetrization these fields satisfy BPS 
condition). We will obtain the energies of quantum BPS fields 
restricting our Hamiltonian  to the neighborhood of the space of BPS 
states and quantizing the restricted Hamiltonian. Let us consider 
the fields of the form 
\begin{equation} \label{con} 
\nabla_{j}=\nabla_{j}^{0} + iq_{j}\cdot {\bf 1} + x_{j} \, ,  
\end{equation} 
\begin{equation} \label{p}
P^{i} = p^{i}\cdot {\bf 1} + \pi^{i}
\end{equation}
where  $x_{i}, \pi^{j} \in End_{T_{\vartheta}}E$, 
${\rm Tr}x_{i} = {\rm Tr}\pi^{j} = 0$. 
For a basic module we can identify $End_{T_{\vartheta}}E$ 
with a noncommutative torus $T_{\tilde \vartheta}$ where 
\begin{equation} \label{varteta}
\tilde \vartheta = (b-a\vartheta)(dim E)^{-1} \,  , \enspace \mbox{and} \enspace 
 a, b \, \mbox{satisfy} \enspace  qb - ap = 1 
\end{equation}
as it was shown in \cite{CR}. Hence, we can 
consider $x_{i}$ and $\pi^{j}$ as functions on a lattice: 
$x_{i} = \sum_{\bf k} x_{i}({\bf k})Z_{\bf k}$, 
$\pi^{j} = \sum_{\bf k} \pi^{j}({\bf k})Z_{\bf k}$ where 
$Z_{\bf k}$ are elements of $T_{\tilde \vartheta}$ satisfying 
\begin{equation} \label{rel1}
Z_{\bf k}Z_{\bf n} = 
exp(2\pi i \tilde \vartheta (k_{2}n_{1} - k_{1}n_{2})) 
Z_{\bf n}Z_{\bf k} \, .  
\end{equation}  
Substituting  expressions (\ref{con}) and (\ref{p}) 
into the Hamiltonian (\ref{hamiltonian}), 
 keeping the terms up to the second order in fluctuations $x_{i}$, 
$p^{i}$  we obtain 
\begin{eqnarray} \label{hfluct} 
&&H_{fluct} =  \frac{R_{0}g_{YM}^{2}dimE}{2R_{1}R_{2}} (p^{i})^{2}R_{i}^{2} +
  \frac{R_{0}}{2g_{YM}^{2}R_{1}R_{2}dimE} (q  + \phi dimE)^{2} + 
\nonumber \\ 
&& +  \frac{R_{0}g_{YM}^{2}dimE}{2R_{1}R_{2}} \sum_{{\bf k}} 
\pi^{i}({\bf k})\pi^{i}(-{\bf k}) R_{i}^{2} + \nonumber \\
&& +  \frac{R_{0}}{2g_{YM}^{2}R_{1}R_{2}dimE}
\sum_{{\bf k}} (k_{1}x_{2}({\bf k}) - k_{2}x_{1}({\bf k}))
(k_{1}x_{2}(-{\bf k}) - k_{2}x_{1}(-{\bf k})) \, . \nonumber \\
&&
\end{eqnarray} 
In the derivation of this formula we used the relation 
\begin{equation} \label{rel2}
[\nabla_{j}, x_{l}] (k_{1}, k_{2}) = \frac{i k_{j}}{dim E} 
x_{l}(k_{1}, k_{2}) \, .  
\end{equation} 
The constraint (\ref{constr}) in the approximation at hand 
now takes the form 
\begin{equation} \label{constr2}
k_{j}\pi^{j}({\bf k}) = 0 \, . 
\end{equation}
In a neighborhood of the space of BPS fields every  field 
satisfying (\ref{constr2}) 
can be transformed by means of a small gauge transformation 
into a field obeying 
\begin{equation} \label{transver}
k_{i}x_{i}({\bf k})R_{i}^{-2} = 0 \, . 
\end{equation}
This means that in our approximation the conditions (\ref{constr2}), 
(\ref{transver}) single out a symplectic manifold $\check {\cal P}$ 
that can be identified with $\tilde {\cal P}={\cal N}/G_{0}$. 
It remains to factorize with respect to large gauge transformations 
to obtain the phase space $\cal P$. The group 
$G^{\large}=G/G_{0}$ of large gauge transformations can be identified 
with the subgroup $G^{mon}$ of $G$ consisting of the elements 
$Z_{\bf k}$ (more precisely, every coset in $G/G_{0}$ has a unique 
representative of the form $Z_{\bf k}$). It is easy to check that 
$\check {\cal P}$ is invariant under the action of $G^{mon}$. 
This observation permits us to identify 
${\cal P}= \tilde {\cal P}/G^{large}$ with $\check {\cal P}/G^{mon}$. 
The Hamiltonian $H$ on $\tilde P$ describes a free motion on a plane and 
an infinite system of harmonic oscillators with frequencies 
 $$
\omega({\bf k}) = \frac{R_{0}}{dimE}
\sqrt{ \frac{k_{1}^{2}}{R_{1}^{2}} + \frac{k_{2}^{2}}{R_{2}^{2}} }
\, . 
$$
More precisely, the fields under consideration can be 
represented in the form 
\begin{equation}
\nabla_{j} = \nabla_{j}^{0} + iq_{j}\cdot {\bf 1} + \sum_{\bf k} 
\mu({\bf k}) x^{\perp }_{j}({\bf k}) 2^{-1/2} 
( a({\bf k}) + a^{*}(\bf{-k}) ) Z_{\bf k}
\end{equation}
\begin{equation}
P^{j} = p^{j}\cdot {\bf 1} + \sum_{\bf k} \pi^{\perp j}({\bf k}) 
\mu({\bf k})^{-1} (dim E)^{-1}2^{-1/2} (a(-{\bf k}) - a^{*}({\bf k}))
Z_{\bf k}
\end{equation}
where $a^{*}({\bf k})$, $a({\bf k})$ are classical counterparts of 
creation and annihilation 
operators obeying the canonical commutation relations, 
$x^{\perp }_{j}({\bf k})$ is a unit vector satisfying  (\ref{transver}), 
$ \pi^{\perp j}({\bf k}) $ is a unit vector satisfying (\ref{constr2}), 
and 
$$
\mu({\bf k}) = \left( \frac{R_{0} g_{YM}^{2}}{R_{1}R_{2}dimE\omega({\bf k})} 
\right)^{1/2} \, . 
$$

 The Hamiltonian now reads as  
\begin{eqnarray} \label{hosc}
H& =&  \frac{R_{0}g_{YM}^{2}dimE}{2R_{1}R_{2}} (p^{i})^{2}R_{i}^{2} +
  \frac{R_{0}}{2g_{YM}^{2}R_{1}R_{2}dimE} (q + \phi dimE)^{2} + 
\nonumber \\ 
&+& \sum_{\bf k} \omega({\bf k}) a^{*}({\bf k}) a({\bf k}) \, . 
\end{eqnarray}
The action of the group $G^{mon}$ on the coordinates 
$q_{j}$, $p^{j}$, $a^{\dagger}({\bf k})$, $a({\bf k})$ can 
be expressed by the formulas
\begin{equation} \label{gaugetr} 
\begin{array}{c} 
q_{j} \mapsto  q_{j} - \frac{n_{j}}{dim E} \\
p^{j} \mapsto p^{j} \\
a({\bf k}) \mapsto exp( i\tilde \vartheta (n_{2}k_{1} - n_{1}k_{2})) a({\bf k})
\\
a^{*} ({\bf k}) \mapsto 
exp(- i\tilde \vartheta (n_{2}k_{1} - n_{1}k_{2})) a^{*}({\bf k})
\end{array}
\end{equation}
These formulas follow immediately from the relations 
(\ref{rel1}), (\ref{rel2}).

The quantization of the system with Hamiltonian  (\ref{hosc}) 
is straightforward. The corresponding space of states is 
 spanned by the wave functions   
\begin{equation} \label{psi}
\Psi_{p_{m}; {\bf k}^{1}, N_{1}; \dots ;{\bf k}^{l}, N_{l}} = 
exp(ip^{m}q_{m}dimE) 
\prod_{j=1}^{l} (a^{\dagger}({\bf k}^{j}))^{N_{j}}|0\rangle
\end{equation} 
where $|0\rangle $ is the oscillators ground state.
Here $a^{\dagger}({\bf k})$ are creation operators and  $p^{i}$ are 
eigenvalues of the quantum operator corresponding to the coordinate $p^{i}$
(which by abuse of notation we denote by the same letter). 
The group $G^{mon}$ acts on the space of states. Under this action the state 
(\ref{psi})   gets multiplied by the 
exponential  factor 
$$
exp\left( i (-n_{j}p^{j} +n_{j}\lambda^{j}+ \tilde \vartheta 
\sum_{j=1}^{l} N_{j}(k^{j}_{1}n_{2} - k^{j}_{2}n_{1}) \right) \, . 
$$ 
where the parameters $\lambda^{j}$ have the meaning of topological 
``theta-angles'' (see Appendix A).  
Thus, the invariance of state vectors under the gauge 
transformations leads to the following quantization law of the 
$p^{i}$-values : 
\begin{eqnarray} \label{pquant} 
p^{1} &=&  e^{1} + \lambda^{1} + 
\tilde \vartheta \sum_{j=1}^{l} N_{j}k_{2}^{j}  
\, , \nonumber \\
p^{2} &=&  e^{2} + \lambda^{2} - 
\tilde \vartheta \sum_{j=1}^{l} N_{j}k_{1}^{j}  
\end{eqnarray}
where $e^{1}$ and $e^{2}$ are integers. 
Substituting this quantization condition into the Hamiltonian (\ref{hosc}) 
we get the energy spectrum 
\begin{eqnarray}\label{BPSspec} 
&&  E = \frac{R_{0}g_{YM}^{2}dimE}{2R_{1}R_{2}}  
(e^{1} + \lambda^{1} + (b-a\vartheta )(dimE)^{-1}\sum_{j=1}^{l} N_{j}k_{2}^{j} 
 )^{2}R_{1}^{2} + \nonumber \\
&& + \frac{R_{0}g_{YM}^{2}dimE}{2R_{1}R_{2}}   (e^{2} + \lambda^{2} -  
 (b-a\vartheta )(dimE)^{-1} \sum_{j=1}^{l} N_{j}k_{1}^{j})^{2} R_{2}^{2} 
  + \nonumber \\ + 
&&\frac{R_{0}}{2g_{YM}^{2}R_{1}R_{2}dimE} (q + \phi dimE)^{2} 
+  \frac{R_{0}}{dimE} \sum_{j=1}^{l} N_{j}
\sqrt{\frac{(k^{j}_{1})^{2}}{R_{1}^{2}} + \frac{(k^{j}_{2})^{2}}{R_{2}^{2}} } 
\nonumber \\
&& 
\end{eqnarray}
Note that to compare this expression with the analogous formulas from 
\cite{Ho}, \cite{BrMor} one should use the relation between the Yang-Mills 
and the M-theory coupling constants:
$$ 
g_{YM} = g_{M}R_{1}R_{2} 
$$  
(see \cite{Taylor} for example). 
Now we would like to show that the expression (\ref{BPSspec}) is invariant 
under $SO(2,2|{\rm \bf Z})$ transformations that 
govern the Morita equivalence. 
It suffices to check that (\ref{BPSspec}) is invariant under the 
transformation $\vartheta \mapsto - \vartheta^{-1}$. 
Using the results of the previous section we write the following 
transformation laws for the quantities constituting expression 
 (\ref{BPSspec}) 
\begin{equation}\label{mor}
\begin{array}{c}
q \mapsto p  ,\,  p\mapsto - q  , \,  
dimE \mapsto \vartheta^{-1}dimE \\
R_{2} \mapsto R_{1}\vartheta , \,  R_{1} \mapsto R_{2}\vartheta  , \,  
R_{0} \mapsto R_{0} \\
\phi \mapsto \phi \vartheta^{2} -  \vartheta , \, 
\lambda^{1} \mapsto \lambda^{2} , \, \lambda^{2} \mapsto -\lambda^{1}\\
e^{1} \mapsto e^{2}  ,\,  e^{2} \mapsto -e^{1} , \, 
k_{1}^{j} \mapsto k_{2}^{j} , \, k_{2}^{j} \mapsto -k_{1}^{j} \\
\tilde \vartheta \mapsto \tilde \vartheta , \, 
g_{YM} \mapsto g_{YM} \vartheta 
\end{array}
\end{equation}
where the change of the coupling constant 
 $g_{YM}$ can be interpreted as an adjustment 
of different trace normalizations on our tori. 
As one can easily check, the expression (\ref{BPSspec}) is invariant 
under transformations (\ref{mor}).


Let us emphasize that the energy 
spectrum (\ref{BPSspec}) is obtained in the assumption  that $E$ is 
a basic module (i.e. in the case when the topological numbers $p$ and 
$q$ are relatively prime). If $D=g.c.d. (p,q) \not= 1$, then the 
factor $D^{2}$ appears in the denominator of the first term in 
(\ref{BPSspec}). This reconciles formula (\ref{BPSspec}) with 
calculations in the commutative case. 


The method used above to obtain the energy spectrum can be 
applied to calculate eigenvalues of a momentum operator. 
Classical momentum functional has the form 
${\rm \bf P}_{i}={\rm Tr}F_{ij}P^{j}$. In the vicinity of a BPS field it 
takes the form
$$
{\rm \bf P}_{i} = q\epsilon_{ij}p^{j} + \sum_{\bf k}\frac{k_{i}}{dimE}
a^{*}({\bf k})a({\bf k}) \, . 
$$ 
The corresponding operator has the following eigenvalues
\begin{equation} \label{momspec}
m_{i} = q\epsilon_{ij}e^{j}  
- a\sum_{j}k_{i}^{j}N_{j}  
\end{equation}
where $a\in {\rm \bf Z}$ is the  integer that enters the 
expression (\ref{varteta}) for $\tilde \theta$ and the parameters $\lambda^{i}$ are assumed 
to be equal to zero.  
Thus, we see that the total momentum is quantized in the usual way 
 (provided $\lambda^{i}=0$). This is not surprising, the integrality 
of eigenvalues is related to the periodicity of the torus. 
One can rewrite the first two terms  of the spectrum (\ref{BPSspec} (contribution of 
``electric charges'')  using the numbers (\ref{momspec}):
\begin{eqnarray} \label{compare}
&&  E = \frac{R_{0}g_{YM}^{2}}{2R_{1}R_{2}dimE}  
(n^{1} + p\lambda^{1} + \vartheta (m_{2}-q\lambda^{1})
 )^{2}R_{1}^{2} + \nonumber \\
&& + \frac{R_{0}g_{YM}^{2}}{2R_{1}R_{2}dimE}   (n^{2} + p\lambda^{2} -\vartheta 
(m_{1}+ q\lambda^{2}) )^{2} R_{2}^{2} 
 \end{eqnarray}
where 
\begin{equation} \label{n}
n^{1} = e^{1}p + b\sum_{j} N_{j}k_{2}^{j} \, , \enspace 
n^{2} = e^{2}p - b\sum_{j} N_{j}k_{1}^{j} \, . 
\end{equation}
(In (\ref{compare}), for simplicity, we set $\lambda^{i}=0$.)
Formula (\ref{compare}) formally matches with the analogous formula of 
\cite{HofVer2} and \cite[v2]{BrMor}  (in particular, the dimension $dimE$ stands in 
the denominator). However, when some of the fluctuations are in the excited state 
they contribute to the energy and the integers $n^{i}$ can not be considered separately 
of the quantum numbers related to fluctuations. When all oscillators representing the 
fluctuations are in the ground state the numbers $n^{j}$ and $m_{i}$ are related by 
the formula $m_{i}p = q\epsilon_{ij}n^{j}$ and thus are not independent. 
One can fix the numbers $m_{i}$ and $n^{j}$ and minimize (\ref{BPSspec}) over all 
$N_{j}$ obeying (\ref{momspec}), (\ref{n}). We obtain 
\begin{eqnarray}\label{1/4BPS} 
&&   E = \frac{R_{0}g_{YM}^{2}}{2R_{1}R_{2}dimE}  
(n^{1} +p\lambda^{1} + \vartheta (m_{2} - q\lambda^{1}) )^{2}R_{1}^{2} + \nonumber \\
&& + \frac{R_{0}g_{YM}^{2}}{2R_{1}R_{2}dimE}(n^{2} +p\lambda^{2} - \vartheta (m_{1} + 
q\lambda^{2}) )^{2} R_{2}^{2} 
  + \nonumber \\ 
&& + \frac{R_{0}}{2g_{YM}^{2}R_{1}R_{2}dimE} (q + \phi dimE)^{2} + \nonumber \\ 
&& + \frac{R_{0}}{dimE}  
\sqrt{\frac{(m_{1}p - n^{2}q)^{2}}{R_{1}^{2}} + \frac{(m_{2}p + n^{1})^{2}q}{R_{2}^{2}} } 
\end{eqnarray}
(When minimizing it is convenient to use the simple fact that a norm of a sum of vectors 
is always larger or equal then the corresponding sum of norms.) 
The last formula agrees with the results of \cite{HofVer2} and contradicts to the 
results of \cite[v2]{BrMor} (the last term is missing in \cite[v2]{BrMor}). 
It is easy to check that (\ref{1/4BPS}) gives energies of 1/4 BPS states; we obtain 
1/2 BPS states when all oscillators are in the ground states 
(i.e.  $m_{i}p = q\epsilon_{ij}n^{j}$).


Let us come back to the consideration of $(1 + d)$-dimensional case. 
We will restrict ourselves to the consideration of modules 
generalizing the basic modules studied in the two-dimensional 
case. We start with the general definition of a basic module. 
We say that a module $E$ over a noncommutative torus $T_{\theta}$ 
is a basic module if the algebra $End_{T_{\theta}}E$ is again a
noncommutative torus $T_{\tilde \theta}$ and the module $E$ is 
equipped with a constant curvature connection $\nabla_{\alpha}$ 
satisfying the condition 
$[\nabla_{\alpha}, \phi ]=\tilde \delta_{\alpha}\phi$ for every 
$\phi \in T_{\tilde \theta}$ (here $\tilde \delta_{1}, \ldots 
\tilde \delta_{d}$ is a basis of the Lie algebra $L_{\tilde \theta}$ 
of infinitesimal automorphisms of $T_{\tilde \theta}$). A 
$T_{\theta}$-module $E$ can be considered as a 
$(T_{\tilde \theta}, T_{\theta})$-bimodule. The conditions 
defining a basic module are equivalent to the condition that 
this bimodule is a complete Morita equivalence bimodule. 
(A complete description of basic modules is given in Appendix D.)
Notice that every Heisenberg\footnote{We use here the term ``Heisenberg 
module'' for the modules described in Sec. 3 of \cite{ASMorita}. }
module $E$ such that $End_{T_{\theta}}E$ 
is a noncommutative torus, is a basic module. (Modules studied in 
\cite{HofVer2} are of this kind.) Let us consider a free 
one-dimensional $T_{\tilde \theta}$-module $T^{1}_{\tilde \theta}$ 
(i.e. $T_{\tilde \theta}$ considered as a right $T_{\tilde \theta}$-module). 
It is easy to check that this module transforms into $E$ by a complete 
Morita equivalence between $T_{\tilde \theta}$ and $T_{\theta}$. 
Using the results of \cite{ASMorita} we arrive at the conclusion that 
we can express energies of the BPS states corresponding to $E$ 
in terms of the BPS states corresponding to $T^{1}_{\tilde \theta}$. 
Let us give the expression of BPS energies in terms of topological 
numbers of the basic module $E$. Without loss of generality we can write 
$$
\mu(E)  = K\cdot exp(-\frac{1}{2}\alpha^{i}Q_{ij}\alpha^{j}) 
$$   
where $K$ is a constant and $Q_{ij}$ is a $d\times d$ matrix. 
It follows from the definition of a basic module 
that there exists a matrix 
\begin{equation} \label{soddz}
\left(
\begin{array}{cc} 
M&N\\
R&S
\end{array} \right) \in SO(d,d|{\rm \bf Z})
\end{equation}
such that $Q=S^{-1}R$. This matrix establishes the Morita equivalence 
between $T_{\theta}$ and $T_{\tilde \theta}$: 
$$
 \tilde \theta = (M\theta + N )(R\theta + S )^{-1} \, . 
$$  
The answer for the energies of 1/2 BPS states is as follows 
\begin{eqnarray} \label{d-BPSspec}
&& E =  \frac{g_{YM}^{2}R_{0}^{2}}{2VdimE}
e^{i} {A_{i}}^{l}g_{lm} {(A^{t})_{j}}^{m} e^{j} + \nonumber \\
&& + \frac{V dimE}{4g_{YM}^{2}} ( (A^{-1}R)_{ij} + \phi_{ij})g^{ik}g^{il} 
( (A^{-1}R)_{kl} + \phi_{kl}) 
\end{eqnarray}
where  $A= R\theta +S$, $dimE = |det A|^{1/2}$, $e^{i}$ are integers.  
It is easy to check that this answer is $SO(d,d|{\rm \bf Z})$-invariant. 
Therefore, it suffices to verify it for a one-dimensional free 
module. The calculation for a free module is based on the same ideas as in 
the two-dimensional case, but technically it is even simpler.  
One should mention, however, that for $d\ge 4$ the group $G^{mon}$ is a proper 
subgroup of the group $G/G_{0} = G^{large}$ of connected components of the group of gauge 
transformations. This remark does not influence the calculation of energies of 1/2 BPS states.

In the $(1+d)$-dimensional case we omitted parameters generalizing the topological 
 ``theta-angles'' $\lambda^{i}$ considered for $d=1$. It is easy to take them into account. 
 

\section*{Appendix A. Geometric quantization. }
It is convenient to use geometric quantization to derive the formulas for 
energies of BPS states. Let us remind the scheme of geometric 
quantization approach. Consider a manifold $\cal X$ with a symplectic 
structure specified by means of a closed form 
$\omega = \frac{1}{2}\omega_{ij}dx^{i}\wedge dx^{j}$. If this form is 
exact ($\omega_{ij} = \partial_{i}\alpha_{j} - \partial_{j}\alpha_{i}$), 
then to every function $F$ on $\cal X$ one can assign an operator 
$\check F$ acting on the space of functions on $\cal X$ by the formula 
$$
\check F \phi = F\phi + \omega^{ij}\frac{\partial F}{\partial x^{j}} 
\nabla_{i} \phi 
$$ 
where  $\omega^{ij}$ is the inverse matrix to $\omega_{ij}$ and 
$\nabla_{i} \phi = \hbar \partial_{i} + \alpha_{i}$ can be considered as a 
covariant derivative with respect to $U(1)$-gauge field having the 
curvature $\omega_{ij}$. It is easy to check that 
$$ 
[\check F, \check G ] = i\hbar (\{ F, G \} )^{\vee}
$$
where $\{ F, G \} = \frac{\partial F}{\partial x^{i}}\omega^{ij} 
\frac{\partial G}{\partial x^{j}} $ is the Poisson bracket. 
However, the correspondence $F\mapsto \check F$ cannot be 
considered as quantization because the operators $\check F$ act 
on the functions depending on coordinates of the phase space $\cal X$. 
The construction of operators $\check F$ is called prequantization. 
In quantization procedure we should construct operators $\hat F$ 
acting on the space of functions depending on $dim {\cal X}/2$ 
variables and obeying $ [\hat F, \hat G ] \approx i\hbar 
\widehat{[F,G]}$ as $\hbar \to 0$. This can be done if we can 
construct an appropriate polarization of $\cal X$ (see \cite{Kirilov}). 
If $\cal X$ is a Kaeler manifold we can use holomorphic polarization 
(i.e. we can define $\hat F$ as an operator acting on holomorphic sections 
of an appropriate holomorphic line bundle over $\cal X$).

Let us  emphasize that the construction of operators $\check F$ in 
the prequantization procedure depends on the choice of the one-form 
$\alpha$ or, better, on the choice of a $U(1)$-gauge field 
having the curvature $\omega$. It is easy to check that replacing 
a $U(1)$-gauge field with a gauge-equivalent field we obtain an 
equivalent prequantization. Different prequantization constructions 
are labeled by the elements of one-dimensional cohomology group 
$H^{1}({\cal X}, {\rm \bf R/Z})$.

We can represent the symplectic manifold $\cal X$ as a quotient space  
$\tilde {\cal X}/\Gamma$ where $\tilde {\cal X}$ is a simply connected space 
(the universal covering of $\cal X$) and the group $\Gamma$, that acts freely 
on $\cal X$, is isomorphic to $\pi_{1}({\cal X})$. Consider a character 
$\chi$ of the group $\Gamma$ and the set $E_{\chi}$ of functions 
on $\tilde {\cal X}$ satisfying the relation 
$F(g x)=\chi(g)F(x)$ for any $g\in \Gamma$. We represent the symplectic 
form $\tilde \omega$ on $\tilde {\cal X}$ as $\tilde \omega = d\alpha$ 
where $\alpha$ is a $\Gamma$-invariant one-form. Then, for every 
$\Gamma$-invariant function on $\tilde {\cal X}$ the prequantization 
construction gives us an operator acting on $E_{\chi}$. Taking into 
account that $\Gamma$-invariant functions on $\tilde {\cal X}$ can be 
identified with functions on $\cal X$ and characters of 
$\Gamma =  \pi_{1}({\cal X})$ can be identified with elements of the 
cohomology group $H^{1}({\cal X}, {\rm \bf R/Z})$, it is easy to check 
that the construction of prequantization in terms of the spaces $E_{\chi}$ 
is equivalent to prequantization on $\cal X$. Using this observation 
we can quantize the theory on $\cal X$ in the following way. 
We quantize $\tilde {\cal X}$ and assume that the group $\Gamma$ of 
symplectomorphisms of $\tilde {\cal X}$ can be lifted to a group of 
unitary transformations acting in the space of wave functions. 
Then it is natural to suppose that the quantum space corresponding to 
$E_{\chi}$ consists of the wave functions satisfying 
$g\psi = \chi(g) \psi$ for all  $ g\in \Gamma$.

Let us consider some examples. We start with a free motion on a circle 
$S^{1}$. Representing $S^{1}$ as $\rm \bf R/Z$ we interpret 
$\pi_{1}(S^{1}) = {\rm \bf Z}$ as a group of translations of 
wave functions $\psi (x) \mapsto \psi (x + n)$ where $n\in {\rm \bf Z}$. 
The space of wave functions corresponding to a character $\chi$ 
consists of functions obeying 
$\psi (x + n) = e^{2\pi i n\theta} \psi (x)$ where $0\le \theta < 1$. 
The spectrum of the Hamiltonian reads 
$$
E_{k} = \frac{(k - \theta)^{2}}{2m} 
$$
where $k\in {\rm \bf Z}$.
Now let us consider a more complicated example where 
$\tilde {\cal X} = {\rm \bf R}^{1}\times {\rm \bf R}^{1}\times 
 ({\rm \bf R}^{2})^{n}$ is a symplectic manifold with coordinates 
$q$, $p$, $a_{1}^{*}, a_{1}, \dots , a_{n}^{*}, a_{n}$ and 
symplectic form 
$$
\omega = dp\wedge dq + i\sum_{j=1}^{n}da_{j}\wedge da^{*}_{j} \, . 
$$
Here $q,p \in {\rm \bf R}^{1}$, $a_{j} \in {\rm \bf C}$. 
Define $\cal X$ as $\tilde {\cal X}/\Gamma$ where $\Gamma$ is 
a cyclic group generated by the transformation 
$$
(q,p,a_{1}^{*},a_{1},\dots , a_{n}^{*}, a_{n} ) \mapsto 
(q+1,p,e^{-i\rho_{1}}a_{1}^{*}, e^{i\rho_{1}}a_{1}, \ldots , 
e^{-i\rho_{n}}a^{*}_{n}, e^{i\rho_{n}a_{n}} ) 
$$
We will consider a Hamiltonian 
\begin{equation} \label{htoy}
H=\frac{p^{2}}{2m} + \sum_{k=1}^{n}\omega_{k}a^{*}_{k}a_{k} 
\end{equation}
on $\tilde {\cal X}$ (a free particle and $n$ independent oscillators). 
This Hamiltonian is $\Gamma$-invariant and therefore generates a 
Hamiltonian on $\cal X$. Wave functions of the theory on $\tilde {\cal X}$ 
can be considered as functions of $q,a_{1}^{*},\ldots , a_{n}^{*}$ 
(we use  the holomorphic representation of oscillator wave 
function). The group $\Gamma$ generates a group of unitary operators 
on these wave functions. The spectrum of  Hamiltonian (\ref{htoy}) reads 
\begin{equation} \label{toyspec}
E = (k- \theta + \sum_{j=1}^{n} N_{j}\rho_{j} )^{2}/2m + 
\sum_{j=1}^{n} \omega_{j}N_{j}  
\end{equation}
where $k\in {\rm \bf Z}$. 

\section*{Appendix B. Spinor representation of $SO(d,d|{\rm \bf Z})$. }

Formula (\ref{Vg}) determines the spinor representation of $SO(d,d|{\rm \bf C})$
as a projective action of  $SO(d,d|{\rm \bf C})$ on the Fock space $F^{*}$. 
This means that the operators $V_{g}$ are defined only up to a constant factor. 
It is possible however to define the spinor representation as a two-valued 
representation of  $SO(d,d|{\rm \bf C})$. This can be done in the following day. 
We introduce a bilinear form on $F^{*}$ defined by the formula 
\begin{equation} \label{bform}
<f,g> = \sum_{k} (-1)^{\epsilon_{k}} f_{k}(\alpha )g_{d-k}(\alpha ) d\alpha_{1} \dots 
d\alpha_{d} \, .
\end{equation}
Here an element of $F^{*}$ is considered as a function of anticommuting variables 
$\alpha_{1}, \dots , \alpha_{d}$ and $f_{k}(\alpha )$ stands for the $k$-th homogeneous
component of $f(\alpha )$. It is easy to check that the numbers $\epsilon_{k}$ can be 
chosen in such a way that linear canonical transformations (operators $V_{g}$) 
preserve the form (\ref{bform}) up to a constant factor. (The verification of this fact can 
be based on the remark that  the  Lie algebra  $so(d,d|{\rm \bf C})$ is 
represented by operators that are quadratic with respect to $a^{k}$, $b_{k}$.) 
This means that we can impose the requirement that  operators $V_{g}$ preserve the 
bilinear form (\ref{bform}). This requirement specifies an operator $V_{g}$ up to a sign; 
in what follows we use this choice of $V_{g}$.

One can define integral elements of the Fock space $F^{*}$ (integral elements of a 
Grassmann algebra) as linear combinations of monomials 
$\alpha_{i_{1}}\cdot  \dots \cdot \alpha_{i_{k}}$ with integer coefficients. An integrality 
preserving operator on 
$F^{*}$ can be defined as a linear operator transforming the set of integral elements into itself, 
or, equivalently, as a linear combination with integer coefficients of monomials composed of 
$a^{k}$, $b_{k}$. For every $g\in SO(d,d|{\rm \bf Z})$ the corresponding operator $V_{g}$ is 
integrality preserving. It is sufficient to verify this fact only for generators of $SO(d,d|{\rm \bf Z})$. 
The group  $SO(d,d|{\rm \bf Z})$ is generated by the transformations 
\begin{equation} 
 (x,y) \mapsto ((A^{t})^{-1}x, A y ) \label{gen1}
\end{equation}
\begin{eqnarray} \label{gen2}
&& (x^{1}, \dots , x^{d}, y_{1},\dots , y_{d}) \mapsto \nonumber \\ 
&& (x^{1},\dots ,y_{i}, \dots ,y_{j},\dots ,x^{d}, y_{1},\dots x^{i},\dots ,  
x^{j}, \dots, y_{d} )
\end{eqnarray}
\begin{equation} \label{gen3}
 (x,y) \mapsto (x, y + Nx) \, .
\end{equation}
Here $(x,y)=(x^{1},\dots ,x^{d},y_{1},\dots ,y_{d})$ is a point of ${\rm \bf R}^{2d}$, 
$A\in SL(d,{\rm \bf Z})$, $N$ is an arbitrary antisymmetric  matrix with integer entries. The  inner 
product in ${\rm \bf R}^{2d}$ is defined by the formula 
$<(x,y), (x',y')> = x^{i}y'_{i} + y_{i}(x')^{i} $. 
The linear canonical transformations corresponding to transformations of the first kind are 
integrality preserving because they are given by the formula $f(\alpha ) \mapsto f((A^{t})^{-1}\alpha )$. 
They preserve the bilinear form (\ref{bform}) due to the fact that the superdeterminant 
of the change of variables $\alpha \mapsto A^{t}\alpha $ is equal to 1. Transformations 
of the third kind generate linear canonical transformations of the form 
$exp(\frac{1}{2}b_{i}N^{ij}b_{j} )$. It is easy to check that these transformations, as well as 
transformations corresponding to generators of the second kind, are integrality preserving and 
preserve 
(\ref{bform}). 
If the operator $V_{g}$ satisfies (\ref{Vg}) and both $V_{g}$ and $V^{-1}_{g}$ are integrality 
preserving 
then $g$ is also integrality preserving , i.e. $g\in  SO(d,d|{\rm \bf Z})$ (we use the fact that a product 
of integrality preserving operators is again an  integrality preserving operator). 
Notice that in this case $V_{g}$ automatically preserves the bilinear form (\ref{bform}).


\section*{Appendix C. Morita equivalent tori.}
In the proof of the results of section 5 of \cite{ASMorita} the spinor representation was 
considered as a projective representation of $ SO(d,d|{\rm \bf Z})$. Therefore, all equations 
were written up to a constant factor. Now we will repeat the proof of these results considering 
the spinor representation as a two-valued representation of $SO(d,d|{\rm \bf C})$ and 
keeping track of all constant factors. We consider completely Morita equivalent 
multidimensional tori $T_{\theta}$ and $T_{\hat \theta}$ and study the relation between 
$\mu = \mu (E)$ and $\hat \mu = \mu (\hat E)$ where $T_{\hat \theta}$-module $\hat E$ 
corresponds to a $T_{\theta}$-module $E$. Using formulas (\ref{8_1/2}) and (\ref{chmu}) 
we obtain 
\begin{equation}
\hat \mu = W\mu \equiv W_{1}W_{2}W_{3}W_{4}\mu 
\end{equation}
where 
\begin{eqnarray}
&&W_{1}f = exp (-\frac{1}{2} b_{k}\hat \theta^{kj} b_{j})f \nonumber \\
&& W_{2}f = exp(  a^{k}\sigma_{kj}a^{j}) f \nonumber \\
&& W_{3}f = \frac{dim\hat E}{dimE} f((A^{t})^{-1}\alpha )\nonumber \\
&& W_{4}f = exp(\frac{1}{2} b_{k}\theta^{kj}b_{j})f \, .
\end{eqnarray}
The operator $W_{1}$ relates $\hat \mu$ and $ch \hat E$, the operator $W_{4}$ relates 
$\mu$ and $ch(E)$ (see (\ref{chmu}). The operator $W_{2}W_{3}$ relates $ch(\hat E)$ and 
$ch(E)$. This relation follows from (\ref{8_1/2}) if we take into account that we should 
identify $L_{\theta}$ and $L_{\hat \theta}$ by means of a linear operator $A$. 
It is clear from the formulas above that the operators $W_{1}, W_{2}, W_{3}, W_{4}$ and 
hence their product that we denoted by $W$ are linear canonical transformations. 
We know that $\hat \mu$ and $\mu$ are integral elements of $F^{*}$, therefore, 
the operator $W$ transforms integral elements of $F^{*}$ into integral elements 
(i.e. $W$ is an integrality preserving operator). The same is true for the inverse operator $W^{-1}$ 
because $\mu$ and $\hat \mu$ are on equal footing. Thus, we can say that the linear 
 canonical transformation $W$ corresponds to an element of  $SO(d,d|{\rm \bf Z})$. 
We proved that $\hat \mu$ and $\mu$ are related by a linear canonical transformation 
corresponding to an element of $SO(d,d|{\rm \bf Z})$. We denote this element by
$$
g = \left( \begin{array}{cc} 
M&N\\
R & S\\
\end{array} \right) \, .
$$ 
 This transformation, as well as transformations $W_{1}$, $W_{2}$, $W_{4}$, 
preserves the bilinear form (\ref{bform}) (for $W$ this follows from integrality of $W$ and 
$W^{-1}$, and for   $W_{1}$, $W_{2}$, $W_{4}$ it can be checked directly). This means that 
$W_{3}$ also preserves (\ref{bform}) and therefore 
\begin{equation}
\frac{dim \hat E}{dimE} = |det(A)|^{-1/2} \, .   
\end{equation}
Going from $W_{1}$, $W_{2}$, $W_{4}$ to the corresponding elements of 
$SO(d,d|{\rm \bf C})$ we obtain 
\begin{eqnarray}
&& 
\left( \begin{array}{cc} 
M&N\\
R & S\\
\end{array} \right) = 
\left( \begin{array}{cc} 
1&\hat \theta\\
0 & 1\\
\end{array} \right) 
\left( \begin{array}{cc} 
1&0\\
-\sigma & 1\\
\end{array} \right)  
\left( \begin{array}{cc} 
(A^{t})^{-1}&0\\
0 & A\\
\end{array} \right)
 \left( \begin{array}{cc} 
1&-\theta\\
0 & 1\\
\end{array} \right) = \nonumber \\ 
&& = \left( \begin{array}{cc} 
(A^{t})^{-1} - \hat \theta \sigma (A^{t})^{-1}&-(A^{t})^{-1}\theta - \hat \theta
( \sigma(A^{t})^{-1} - A)\\
-\sigma(A^{t})^{-1} & \sigma  (A^{t})^{-1}\theta + A \\
\end{array} \right) \, .   
\end{eqnarray} 
From the last formula one readily derives formulas (\ref{A}), (\ref{sigma}) and 
(\ref{newteta}).


\section*{Appendix D. Modules with constant curvature connection.} 
One can give a complete description of modules over noncommutative tori $T_{\theta}$ that 
can be equipped with a constant curvature connection (\cite{AstSchw}). 
In this appendix we will give this description for the case when the matrix $\theta$ is 
irrational and, moreover, any linear combination of its entries is irrational provided all 
coefficients are integers.

For every element $x\in F^{*}$ we can define a subset $T_{x}$ of ${\rm \bf R}^{2d}$ 
consisting of vectors $(u,v)\in {\rm \bf R}^{2d}$  that obey 
$$ 
(u^{i}\frac{\partial}{\partial \alpha^{i}} + v_{i}\alpha^{i} ) x = 0 \, . 
$$
It is easy to check that $T_{x}$ is an isotropic linear subspace of ${\rm \bf R}^{2d}$.
(We equip ${\rm \bf R}^{2d}$ with the same bilinear product as in Appendix B). 
We will say that an even element $x\in F^{*}$ is a generalized quadratic exponent (GQE) 
if $T_{x}$ is a maximal isotropic subspace of ${\rm \bf R}^{2d}$ (i.e. $dimT_{x} = d$). 
One can say that GQE $x$ satisfies 
\begin{equation} \label{GQE}
(U_{i}^{j} \frac{\partial}{\partial \alpha^{i}} + V_{ij}\alpha^{j} )x=0
\end{equation}
where $(U,V)=(U_{i}^{j},V_{ij})$ is a $d\times 2d$ matrix of rank $d$. 
Let us represent $x$ in the form 
$$
x=N + \frac{1}{2}\alpha^{i}m_{ij}\alpha^{j} + \dots = N + \frac{1}{2}\alpha M \alpha + \dots 
$$
where the omitted terms have higher order with respect to $\alpha$.
It is easy to check that $VN + UM = 0$.
 If the matrix $U$ is nondegenerate, we can express $x$ in terms of $N$ and $M$ solving 
the differential equation (\ref{GQE}). We obtain 
$$
x=N \cdot exp(-\frac{1}{2}\alpha N^{-1}M\alpha )
$$
where $M=-U^{-1}VN$. We see that $x$ is a quadratic exponent; it is easy to check that the 
set of quadratic exponents is a dense subset of the set of all GQE. Now we can formulate 
the following theorem. \\
{\it A module $E$ over a noncommutative torus $T_{\theta}$ can be equipped with a constant curvature 
connection iff $\mu (E)$ is a GQE. The module $E$ is basic iff $\mu (E)$ is a GQE without  nontrivial 
divisors (i.e. $\mu (E)$ cannot be represented in the form $\mu (E)=D\nu $ where 
$D\in {\rm \bf Z}$, $D>1$, and $\nu$ is an integral element of $F^{*}$). }\\
To prove this theorem we notice first of all that the set of all GQE is invariant 
under projective action of $SO(d,d|{\rm \bf Z})$ in $F^{*}$. It follows immediately from 
(\ref{Chern}) that if module $E$ admits a constant curvature connection, then $ch(E)$ 
is a quadratic exponent. Taking into account that $\mu (E)$ is related to $ch(E)$ by 
means of a linear canonical transformation, we obtain that $\mu (E)$ is a GQE.

Now we should assume that $\mu (E)$ is a GQE and prove that $E$ can be equipped with a 
constant curvature connection. It is sufficient to find such an element 
$g\in SO(d,d|{\rm \bf Z})$ that 
\begin{equation}\label{const}
V_{g}\mu (E) = const. 
\end{equation}
Then we can use this element $g$ to transform $T_{\theta}$ into a completely Morita equivalent 
torus $T_{\hat \theta}$. The module $\hat E$ over $T_{\hat \theta}$ corresponding to 
the module $E$ by means of this construction is a free module. 
The results of \cite{ASMorita} give a correspondence between constant curvature connections in 
$E$ and in $\hat E$. A free module has a zero curvature connection. Therefore, 
(\ref{const}) implies the existence of constant curvature connection on $E$. 
As the first step in the construction of $g$ satisfying (\ref{const}) we construct  
such an element $h\in SO(d,d|{\rm \bf Z})$ that $V_{h}\mu (E)$ is a quadratic exponent. 
This is easy to do applying generators (\ref{gen2}) of $SO(d,d|{\rm \bf Z})$ and 
taking into account that the rank of the matrix $(U,V)$ is equal to $d$. The next step is 
to simplify the quadratic exponent by means of transformations (\ref{gen1}). Our consideration 
will be similar to the arguments applied in \cite{HofVer2} to a different problem. It is known 
(see \cite{Igusa}) that  antisymmetric bilinear form with integer entries can be reduced to a block-diagonal 
form with $2\times 2$ blocks by means of $SL(d,{\rm \bf Z})$ transformations. 
This means that we can restrict ourselves to the case when $\mu (E)$ is an integral element 
of $F^{*}$ having the form 
$$
\mu (E)=N\cdot exp(\frac{1}{N}(m_{1}\alpha^{1}\alpha^{2} + m_{2}\alpha_{3}\alpha_{4} + 
\dots + m_{k}\alpha^{2k-1}\alpha^{2k}))
$$
where $m_{1},\dots ,m_{k}$ are non-zero integers. 
 Now we will use an element of $SO(d,d|{\rm \bf Z})$ that can be 
represented by means of block-diagonal transformation with blocks 
$$
\left(\begin{array}{c}\tilde x^{2i-1}\\
\tilde x^{2i} \\ \tilde y_{2i-1} \\ \tilde y_{2i} \end{array}\right) = 
\left( 
\begin{array}{cccc}
N/n_{i} & 0 &0& k_{i}\\
0& N/n_{i} &-k_{i}&0\\
0& -m_{i}/n_{i} & l_{i} &0 \\
m_{i}/n_{i}&0&0&l_{i}  
\end{array}
\right) \cdot 
\left(\begin{array}{c} x^{2i-1}\\
 x^{2i} \\  y_{2i-1} \\  y_{2i} \end{array}\right)
$$
where $n_{i} = g.c.d.(m_{i},N)$, and   $k_{i}$, $l_{i}$ are integers 
satisfying $l_{i}N - m_{i}k_{i} = n_{i}$. 
It is easy to check that after this transformation (\ref{const}) is satisfied.

Notice that  to consider Morita equivalence with a given $g\in SO(d,d|{\rm \bf Z})$ we
should require that the corresponding fractional linear transformation of $\theta$ is 
well defined. This condition is satisfied due to the assumptions imposed on $\theta$. 

\begin{center} {\bf Acknowledgments}  \end{center}

We are indebted to A.~Astashkevich, A.~Connes, C.~Hofman, B.~Pioline, 
M.~Rieffel, and B.~Zumino for useful discussions.


 \end{document}